\documentclass[aps,prl,twocolumn,floats]{revtex4}
\usepackage{amsmath,bm,epsfig}

\def\Fbox#1{\vskip1ex\hbox to 8.5cm{\hfil\fboxsep0.3cm\fbox{%
  \parbox{8.0cm}{#1}}\hfil}\vskip1ex\noindent}  
\let \nn  \nonumber
\newcommand{\br}{\\ \nn}

\let\*\cdot
\def\<{\left\langle} \def\>{\right\rangle} \def\({\left(} \def\){\right)}
\let\p\partial \let\~\widetilde \let\^\widehat \def\ort#1{\^{\bf{#1}}}
\def\Trans{^{\scriptscriptstyle{\rm T}}} \def\x{\ort x} \def\y{\ort y}
\def\z{\ort z} \def\bn{\bm\nabla} \def\1{\bm1} \def\Tr{{\rm Tr}}
\def\pp {\perp} \def\pl {\parallel}
\newcommand{\B}[1]{{\bm{#1}}}
\newcommand{\C}[1]{{\mathcal{#1}}}    
\newcommand{\BC}[1]{\bm{\mathcal{#1}}}
\newcommand{\F}[1]{{\mathfrak{#1}}}
\newcommand{\BF}[1]{{\bm{\F {#1}}}}
\def\BE{\begin{equation}}\def\EE{\end{equation}}
\def\BEA{\begin{eqnarray}}\def\EEA{\end{eqnarray}}
\def\BSE{\begin{subequations}}\def\ESE{\end{subequations}}

\renewcommand{\sb}[1]{_{\text {#1}}}  
\renewcommand{\sp}[1]{^{\text {#1}}}  
\newcommand{\Sb}[1]{_{_{\text {#1}}}} 
\newcommand{\Sp}[1]{^{^{\text {#1}}}} 
  \def\Sb#1{_{\scriptscriptstyle\rm{#1}}}

\newcommand{\Ref}[1]{(\ref{#1})}
\newcommand{\REF}[1]{Eq.~(\ref{#1})}

\renewcommand{\a}{\alpha}\renewcommand{\b}{\beta}\newcommand{\g}{\gamma}
\newcommand{\G} {\Gamma}\renewcommand{\d}{\delta}
\newcommand{\D}{\Delta}\newcommand{\e}{\epsilon}\newcommand{\ve}{\varepsilon}
\newcommand{\E}{\Epsilon}\renewcommand{\o}{\omega} \renewcommand{\O}{\Omega}
\renewcommand{\L}{\Lambda}
\renewcommand{\t}{\tau}
\def\R{\bm R}
\def\s{\sigma}
\def\S{\Sigma}\def\R{\mathcal R}
\let\p\partial \def\bn{\bm\nabla}
\def\x{\ort x} \def\y{\ort y}  \def\z{\ort z}
\def\E{\mathcal E}  \def\P{\mathcal P}
\def\vt {v\Sb T }
\def\ut {u\Sb T } \def\uz  {u \! _{_ \perp }}
\def\Wd {W^\ddag} \def\Vd {V^\ddag}
\def\kd {k^\ddag}\def\gd {g^\ddag} \def\kpd {\kd\sb p}
\def\ld {\ell^\ddag}
\def\Sd {S^\ddag} \def\ud {u^\ddag}
\def\utd {u\Sb T ^\ddag}\def\uzd  {{\utd}_z}
\def\yp {y^+}\def\zp {z^+}
\def\yd {y^\ddag}\def\zd {z^\ddag}
\def\Wp {W^+} \def\Vp{V^+}
  \def\Gd {\Gamma^\ddag}\def\Usd   {U\Sb S ^\ddag}
 \def \vK {von-K\'arm\'an~}
\def\V{\bm V}
\def\w{\bm w}
\def\W{\bm W}
\def\lG {\overline{G }}

\let \= \equiv

\def\hf{\frac{1}{2}}

 \let\*\cdot
 \def\sb#1{_{\rm{#1}}}

 \def\({\left(} \def\){\right)}
 \def \[ {\left [} \def \] {\right ]}
   \def\Sp#1{^{\scriptscriptstyle\rm{#1}}}
    \let\^\widehat
  \let\-\overline

  \def\REF#1{Eq.~\Ref{#1}}
   \def\Ref#1{(\ref{#1})}  \def\<{\left\langle}
   \def\>{\right\rangle}
   \def\ort#1{\^{\bf{#1}}}

   \let\~\widetilde
\def\Re{\ensuremath{{\C R}\mkern-3.1mu e}}
\def\De{\ensuremath{{\C D}\mkern-3.1mu e}}
\def\Rid{\mbox{Ri}^\ddag}\def\Rip{\mbox{Ri}^+}
\begin{document}
\title{Analytic Model of the Universal Structure of
Turbulent Boundary Layers}
\author{Victor S. L'vov,  Itamar Procaccia and Oleksii Rudenko}
\affiliation{The Department of Chemical Physics, The Weizmann
Institute of Science, Rehovot 76100, Israel}

\begin{abstract}
Turbulent boundary layers exhibit a universal structure which
nevertheless is rather complex, being composed of a viscous
sub-layer, a buffer zone, and a turbulent log-law region. In this
letter we present a simple analytic model of turbulent boundary
layers which culminates in explicit formulae for the profiles of the
mean velocity, the kinetic energy and the Reynolds stress as a
function of the distance from the wall. The resulting profiles are
in close quantitative agreement with measurements over the entire
structure of the boundary layer, without any need of re-fitting in
the different zones.
\end{abstract}
\maketitle

\textbf{Introudction}: Theoretical physicists tend to consider
turbulence in the context of  the  idealized model of isotropic and
homogenous fluid flows at large Reynolds numbers. In part this is
due to the apparent existence of universal, anomalous scaling
exponents which characterize correlation and structure functions in
fully developed turbulent flows. It is also easier to disregard the
effects of walls which introduce essential anisotropies and
inhomogeneities. Needless to say, all realistic turbulent flows are
neither homogeneous nor isotropic. A problem of extreme interest for
both technological applications and from the point of view of basic
science is ``wall-bounded" turbulence, with the theoretical model of
a flat infinite wall playing a key role. This problem presents also
fascinating universal features, see Fig.1, but traditionally it was
more popular in the engineering rather than in the physics
community. There are fascinating open problems in wall-bounded
turbulence. The present Letter attempts at finding a simple model
that affords an analytic calculation of the universal profiles of
the mean velocity, the turbulent energy and the Reynolds stress as a
function of the distance from the wall.
\begin{figure}
  \centering
   \includegraphics[width=0.47   \textwidth]{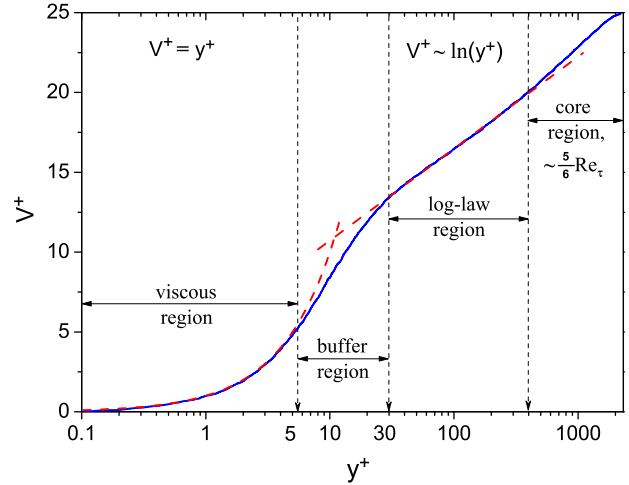}
  \caption{\label{f:0} A sketch of the characteristic regions
  appearing in a turbulent boundary layer.  The continuous line
  represents the mean velocity profile
    $V^+(y^+)$ [in the wall units, \REF{y+}]
  taken from  direct numerical simulation at  Re$_\tau$
  = 2320, \cite{Re2320}. The "viscous region" ends when the linear
  law $V^+\!\! = y^+$ begins to deviate from the continuous line at
  $y^+\approx 6$. The "log-law region" ends approximately at $1/6$
  of the channel hight, i.e. $y^+\!\!\simeq \textrm{Re}_\tau/6$. At
  this point and deeper towards the channel cenerline the
  dimensionless momentum flux to the wall, $\mathcal{P}^+\! = 1 -
  {y^+}\!/\,{\mbox{Re}_\tau}$ (see~\REF{2FSol-inv}), deviates
  appreciable from unity.}
\end{figure}

The theory that we construct begins with the equations of fluid
mechanics and focuses on the momentum and energy fluxes using the
conservation laws for these quantities as a guidance for developing
an appropriate model. In gross substance this approach is not new,
and indeed a number of ingredients are borrowed from the literature.
The model that we end up with is however improved compared to
previous results in the sense that it provides analytic predictions
for the above-mentioned profiles in the entire boundary layer,
without re-fitting in the different zones. It should be stressed at
this point that one cannot expect a universal model to apply for all
turbulent boundary values problem. For instance, the understanding
of drag reduction by additives calls for a slightly modified model
that stresses the effects of the additives. The present contribution
offers what we consider a simplest model that is constructed to best
describe Newtonian wall-bounded turbulence with enough richness to
capture all the essential universal profiles of the quantities of
interest.  One advantage of the model that will be demonstrated in
future publications is that it can be naturally generalized to
describe stratified turbulent boundary layers caused by, say,
temperature gradients,  heavy particles, etc, with applications to
interesting problems like sand storms in deserts,  snow fall in a
windy days or water
flow in a heavily silt-laden rivers.

\textbf{Equations and definitions}: The starting point are the
Navier-Stokes equations for an incompressible fluid velocity $\bm
U(\bm r,t)$,
\begin{equation}
 \frac{\partial \bm U}{\partial t} +\(\bm U\cdot \bm \nabla\)\bm
U =-\frac{\bm \nabla p}{\rho}  +\nu \Delta \bm U \,, \quad \bm \nabla\cdot \bm U = 0\,,
\label{NSE1}
\end{equation}
where $\rho$ is the fluid density, $p=p(\bm r,t)$ -- the pressure
and $\nu$ is the kinematic viscosity. We follow the standard
strategy of Reynolds  considering the velocity as a sum of its
average (over time) and a fluctuating part:
\begin{equation}
\bm U(\bm r,t) = \bm{V}(\bm r)  + \bm u(\bm r,t) \ ,
\quad \bm V(\bm r) \equiv \langle \bm U(\bm r,t) \rangle \ . \end{equation}
We also introduce the conventional
\emph{viscous scale} $\ell_\tau$ and \emph{friction velocity}
$u_\tau$ 
\BE
u_\tau \= \sqrt{ {P(0)}/{\rho}}\,, \quad \ell_\tau \=
 {\nu}/{u_\tau}\,, 
\EE
where $P(0)$ is the wall shear stress -- the flux of the mechanical
momentum at the wall. This quantity depends on the type of turbulent
boundary layer (pressure driven or driven by the upper boundary,
etc). The velocity and the distance from the wall are then measured
in \textit{wall units}
\BE
\label{y+}
\B U^+ \!\!\equiv  {\B U}/u_\tau\,,~    y^+ \!\!\equiv
 {y}/{\ell_\tau}\ , ~ p^+\!\!\equiv p/(\rho u^2_\tau)\ ,
  ~ \B \nabla^+\!\!\equiv \ell_\tau \B \nabla.\EE
In wall units the Navier-Stokes equation is now dimensionless, and
the stationary version takes the form
\begin{equation}
 \(\bm U^+\cdot \bm \nabla^+\)\bm U^+ =-\bm \nabla^+ p^+  +\Delta^+ \bm U^+ \
 .
\label{NSE2}
\end{equation}
The averaged equation for the mean velocity looks like
\BE\label{MeanV3D}
 \(\bm V^+\cdot \bm \nabla^+\)\bm V^+  = \Delta^+ \B V^+ -\langle \B u^+
\cdot \B \nabla^+ \B u^+\rangle    -\B \nabla^+ \langle p^+\rangle\ .
\EE

Besides the mean velocity one needs to consider correlation functions. It turns
out that  important  features of wall bounded turbulence, like the mean
velocity profile, thermal conductivity, turbulent transport of
matter, etc., are determined by velocity fluctuations on relatively large
scales, and the statistics of the latter do not deviate too much from
Gaussian statistics. Thus an economic description of these
features can be reached on a level of second order correlation
functions. Therefore,  in addition to the mean velocity profile,  we will be
interested in the detailed description of  two additional
quantities, the turbulent energy and the Reynolds stress tensor:
\begin{equation}
K^+ \equiv  \langle  |\B u^+|^2\rangle /2  \ , \quad W^+_{ij} \=
\<u^+_i u^+_j\>\ . \EE

\textbf{Balance equations for the Reynolds stress}: Subtracting Eq.
(\ref{MeanV3D}) from Eq. (\ref{NSE2}), multiplying by a fluctuating
velocity component $u^+_j$ and averaging, results in the equation
satisfied by the Reynolds stress: \BE\label{corr} \(\bm V^+\cdot \bm
\nabla^+\) W_{ij}^+ = P^+_{ij} +\C \R^+_{ij}
-\varepsilon^+_{ij}+\partial_kT^+_{ijk}\,,
 \EE 
 where the tensors of
the energy production, $P^+_{ij} $, of the pressure-rate-of-strain,
$\C R^+_{ij} $ and of the Reynolds-stress dissipation
$\varepsilon^+_{ij} $,  are 
\BEA \label{defs1} 
  P^+_{ij}&\!\!\=\!\! &-
W^+_{ik} \partial_k V^+_j
+W^+_{jk} \partial_k{V^+_i} ,
 \br
 \C R_{ij}^+&\!\!\=\!\!& \left\langle \~p\,^+ \(
\partial_j{u^+_i}+\partial_i {
u^+_j} \)\right\rangle, \quad \varepsilon^+_{ij}
\= 2\langle \partial_k {u^+_i}\,
\partial_k{u^+_j} \rangle\ ,   
\EEA
and $\~p \= p -\<p\>$ denotes the pressure fluctuations. The last term
$T^+_{ijk}$ presents spatial energy fluxes. We will neglect it
throughout the  turbulent boundary layer; it is indeed small in the
log-layer, but comparable to other terms in the buffer and viscous sub-layers.
The model will be constructed such as to compensate for this neglect in
those regions where the term is significant. The bonus of neglecting
this term is enormous since this keeps the theory {\em local}, without
partial derivatives.

The modeling of the various terms appearing in Eq. (\ref{corr}) has
attracted considerable attention over the years, and we only briefly
summarize how this is done. The Poisson's equation for the
fluctuating pressure  follows from the equation of the fluctuating
part of the velocity field, $\bm u$, which is obtained by
subtracting  \REF{MeanV3D} from  \Ref{NSE2}:
\BSE \label{pr1} \BE \label{pr1a}
\Delta^+\~p\,^+ = -\nabla^+_i \nabla^+_j \( u^+_i u^+_j -\<u^+_i u^+_j\>
 +V^+_i u^+_j +V^+_j u^+_i\). 
\EE
 The homogeneous solution of this equation is
 responsible for sound, a phenomenon of very little consequence for
 turbulent dynamics at low Mach numbers. The inhomogeneous solution includes
 two parts, $ \~{p}\,^+ = \~p_{uu}^{\,+} +\~p^{\,+}_u$: 
  \BE \label{pr1b}
 \~p^{\,+}_{uu}\propto u^+_i u^+_j -\<u^+_i u^+_j\>\,,
\quad   \~p^{\,+}_u \propto  V^+_i u^+_j+ V^+_j u^+_i\ .
 \EE \ESE 
 Correspondingly the pressure-rate-of-strain tensor~\Ref{defs1}
 consist of two terms: 
\BSE\label{ROF}
\BE\label{rof} 
\C R_{ij}^+= {R_{ij}\Sp{\, RI}}^+ +{R_{ij}\Sp {\,IP}}^+.
\EE 
The first of these is known as the ``Return-to-Isotropy" tensor,
${R_{ij}\Sp{\, RI}}^+$, that depends on the tripple-velocity
correlator $\<u_iu_ju_k\>$. Its evaluation in terms of the objects
of the theory calls for a closure, and following time-honored
tradition~\cite{Pope} we adopt for it the simple Rota form
\BE\label{RIa} 
{R_{ii}\Sp{\, RI}}^+  \simeq   -\g\Sb {RI}\(3\,W^+_{ii} -W^+\) ,
\quad W^+ \= \Tr\{ W^+_{ij}\}\,, 
\EE 
in which  $\g\Sb{RI}$ is some characteristic nonlinear frequency
that will be specified later. The tensor ${R_{ij}\Sp{\, RI}}^+$ is
traceless and therefore the frequency $\g\Sb{RI}$ must be the same for
all diagonal components of ${R_{ii}\Sp{\, RI}}^+$. There is no
reason however to assume that off-diagonal terms have the same frequency.
Therefore, following~\cite{MMI}, we assert that 
\BE\label{RIb}
{R_{ij}\Sp{\, RI}}^+  \simeq   -
3\,\~\g\Sb {RI} W_{ij}  \,, \quad i\ne j\,,
\EE
with, generally speaking, $\~\g\Sb {RI}\ne \g\Sb {RI}$.

The traceless ``Isotropization-of-Production" tensor, ${R_{ij}\Sp
{\,IP}}^+$ has a structure that is very similar to the production
tensor $\C P_{ij}^+$, \REF{defs1},  and thus traditionally it is
modeled in terms of $\C P_{ij}^+$~\cite{Pope}:
 \BE \label{IP}
{R_{ij}\Sp {\,IP}}^+ \simeq  -C\Sb {IP}\( 3\,\C \P^+_{ij} -\C
P^+\delta_{ij}\)\,,  \  \C P^+\= \mbox{Tr} \,\{ \C \P^+_{ij}\}\ .
 \EE
 \ESE

The dissipation tensor  $\ve^+_{ij}$ is estimated differently far
from the wall and near it. Far  from the wall and for large Reynolds
numbers the turbulent flow can be considered approximately
isotropic. Therefore, the tensor $\ve^+_{ij}$ should be
approximately diagonal,
\BSE\label{distau1}
\BE\label{distau1a}
\ve^+_{ij}=\gamma^+\,W^+\, \d_{ij}\ .
\EE
Under stationary conditions the rate of turbulent kinetic
energy dissipation is equal to the energy input at the outer scale,
estimated as $\< u_i u_j u_k \>/ \ell$ where   the outer scale of
turbulence $\ell$ is estimated as the distance to the wall $y$.
Therefore, the natural estimate of $\g ^+$ involves the
tripple-velocity correlator,  
\BE\label{distau1b} 
\gamma^+\sim\frac{ \< u u u \>^+}{ y \< uu \>^+}  \ \Rightarrow \
\gamma^+ \! = b\,\frac{\sqrt{W^+}}{y^+}\ . 
\EE 
\ESE 
Similarly, we can estimate  the Return-to-Isotropy frequencies $\g
^+ \Sb{RI}$ and $ \~{\g}^+\Sb{RI}$ in Eqs.~\Ref{RIa} and \Ref{RIb}.
Having in mind that the precise structure (tensorial contraction,
etc.) of the equation for $\g^+$  is different from that of the
 equations for  $\g^+ \Sb{RI}$ and $ \~{\g}^+\Sb{RI}$, we should involve different numerical
prefactors: 
\BE \gamma^+ \Sb{RI} = b \Sb{RI}\, \sqrt{W^+}/y^+ \,, \quad
\~\gamma^+ \Sb{RI} = \~b \Sb{RI}\, \sqrt{W^+}/y^+ \ .
\EE 
Close to the wall, in the viscous sub-layer $y^+\le 30$, the
estimates change due to the direct viscous contribution of the
largest eddies at that distance, whose size is of the order of the
distance $y^+$ itself.  For these eddies we can estimate the $\nabla^2$
operator as $(\~ a/y)^2$, where $\~a$ is a new  fitting constant. For
simplicity we account for this contribution only in off-diagonal
terms, which do not include the nonlinear part (see \REF{distau1a}). In this way we will have:
\BSE \BE \label{distau} 
\ve^+_{ij} = \~\Gamma^+\, W_{ij}^+\,, \quad i\ne j\,, 
\EE  
with $\~ \Gamma ^+\!\!=(\~a/y)^2 $.  Here we have to recall that we
neglected the spatial energy transfer term, which plays an important
role in the viscous sub-layer. Since it has an opposite sign to the viscous
dissipation, we can take its influence into account by
suppressing the direct viscous dissipation by a a function of $W/W_*$ (here
$W_*$ is the asymptotical value of $W$ in the lag-low region):
\BE\label{tG} ~\Gamma^+ = \( \~a/y^+ \)^2 \sqrt{  {W^+}/{ W^+_*} } \
.\EE \ESE  
Our choice of a square-root function is dictated by the simplicity of the
analytical treatment of the resulting algebraic model.  A-posteriori,
the implicit  accounting for the energy flux in this particular way
 is supported by the good agreement of
the model prediction for the energy profile in both the viscous and
the buffer layers with the DNS data without additional fitting
parameters, see the insert in Fig.~\ref{f:2}.

 \textbf{Plane geometry and the balance of momentum:} For plane
geometry the mean velocity is oriented in the (stream-wise) $\x$
direction and depends only on the vertical (wall-normal) coordinate
$y$: $\bm V = V(y)\,\x$. For such flows all the averages are
functions of $y^+$ only. An interesting special example is a channel
flow of height $2L$ and infinite extent in the span-wise direction.
Due to the symmetry in the span-wise direction $z\to -z$,
$W^+_{xz}=W^+_{zz}=0$. From Eq. \Ref{MeanV3D} for $V$ (integrated
over $y^+$) one gets the exact balance equation for the
mechanical-momentum 
 \BEA\label{MeanMom}  P^+(y^+)\!\! &=& \! S^+\! -W^+_{xy} \,, \quad
\mbox{where}\\ \nn 
S^+(y)\!\! &\=&\! \frac{d V^+\!}{d y^+\!}\,, \  P^+(y^+) \= 1 -
\frac{y^+}{\textrm{Re}_{\tau}\!}\,,\  \textrm{Re}_{\tau}\! \= \frac{L
\,u_\tau}{\nu}\ .
 \EEA
The flat geometry also simplifies the production term defined in
\REF{defs1}:
\BE\label{2DProduction} \P^+_{ij} = -S^+\(W^+_{iy}\,\delta_{jx}
+W^+_{jy}\,\delta_{ix}\) . 
\EE 

\textbf{Final set of equations:} Substituting everything into
\REF{corr} we get the following set of model equations: 
\BSE\label{MainEqns} 
 \BEA \label{xx}\!\!\!\!\!\!\!\! 3\gamma^+ \Sb{RI} W^+_{xx} \!\! &\!=\!&
\!\! \(\gamma^+ \Sb{RI} -\gamma^+\)
 W^+ \!- \!2\(1 -2\,C\Sb{IP}\)S^+W^+_{xy},~~~~~~\,\\ \label{yy}
3\gamma^+ \Sb{RI} W^+_{yy} \!\! &=& \!\! \(\gamma^+ \Sb{RI} -
\gamma^+\) W^+ -2\,C\Sb{IP} S^+W^+_{xy} \,,  \\ \label{zz}
3\gamma^+
\Sb{RI} W^+_{zz} \!\! &=& \!\! \(\gamma^+ \Sb{RI} -\gamma^+\)
W^+ -2\,C\Sb{IP} S^+W^+_{xy} \,, \\
0 \!\! &=& \!\!  \big (\~\Gamma^+\!  +  3\~\gamma^+ \Sb{RI}\big )
W^+_{xy} +\(1 -3\,C\Sb{IP}\)
 S^+W^+_{yy}\,.
\EEA\ESE
Summing up Eqs.~\Ref{xx}, \Ref{yy} and  \Ref{zz} [or, equivalently,
taking the trace of \REF{corr}] one gets 
\BE\label{MainTrace}
3\,\gamma^+ W^+ =  -2\,S^+ W^+_{xy} \ . \EE

\textbf{Choice of the log-law   parameters $b$, $b\Sb{RI}$ and $\~b
\Sb {RI}$: }
 At this point we can fit three  $b$-parameters
responsible for the solution in the asymptotical  log-law region
$30 < y^+ < \textrm{Re}_\tau/6$ for sufficiently large Reynolds numbers. In this
region $\~\Gamma^+$  can be neglected, and solution of
Eqs.~\Ref{MainEqns} takes the form  
\BSE\label{FormalSol}
\BEA\label{FormalSol-1}
W^+_{xx} &=& \frac{\(1-2\,C\Sb{IP}\)
b +b\Sb{RI}}{b+3\,b\Sb{RI}}\,W^+_{\ast} \,,  \\
\label{FormalSol-2}
W^+_{yy} &=& W^+_{zz} = \frac{C\Sb{IP}b +b\Sb{RI}}{b
+3\,b\Sb{RI}}\,W^+_{\ast}\,,
 \\
\label{FormalSol-3}
W^+_{xy} &=& - b \, \kappa  \(W^+_{\ast} \)^{3/2}=-1\,, \\
\label{FormalSol-4}
S^+_{\ast}& = &  \frac1 {\kappa y^+}\,,\quad \mbox {where}\\
\kappa  &=&\sqrt{ \frac{2\(1 -3\,C\Sb{IP}\)\(C\Sb{IP}b +b
\Sb{RI}\)}{3\, W^+_{\ast} \,b \, \~ b\Sb{RI}\(b +3\,b\Sb{RI}\)}} \ .
\EEA
\ESE
 To determine $b$-parameters, we used the following
data:\\

1.  The numerical values of $W^+_{\ast} = 7.8$ and von-Karman
constant $\kappa = 0.405$ can be  taken from the direct numerical
simulations (DNS) \cite{DNS};

2.  The detailed analysis of  experimental, DNS and
large-eddy-simulation data, made in Ref.~\cite{MMI}, yields the
conclusion that with a good accuracy one can take 
\BE\label{ani}
 W^+_{xx }= 2\, W^+_{yy }=2\, W^+_{zz } = {W^+_{\ast}}/{2}\, ; \EE

3. The suggested in the literature (see, e.g.~\cite{Pope}) value of
 $C\Sb{IP}$ is $\frac 15 $. Our analysis showed that all profiles are very insensitive to a particular choice of
 $C\Sb{IP}$ around value 0.2. Therefore for simplicity we take
 \BSE \label{all-b}
   \BE\label{CIP} 
C\Sb{IP}= 1/5\ . 
\EE
\vskip -0.2cm
All  this knowledge enables us to find   $b$, $b \Sb{RI}$ and $\~b
\Sb{RI}$: \BEA
\label{b}
b &=& \frac{2}{3\,\kappa\, {W^+_{\ast}}^{3/2}} \approx 0.075 \,,
\\ 
\label{bRI}
b \Sb{RI} &=& 4\(1 -3\,C\Sb{IP}\)b \approx 0.121 \,,\\ 
\label{tbRI}
\~b \Sb{RI} &=& \frac{\(1 -3\,C\Sb{IP}\)\sqrt{W^+_{\ast}}}{12\,\kappa}
 \approx 0.230 \ .
\EEA
For future purpose  we  introduce a parameter
\BE\label{bbar}
\~b\= \frac{\~b \Sb{RI}}{1 -3C\Sb{IP}}
 =\frac{\sqrt{W^+_{\ast}}}{12\,\kappa} = 0.575\ .
\EE \ESE

\textbf{General solution:} With the chosen parameters~\Ref{all-b} the
formal solution of the system \Ref{MeanMom}, \Ref{MainEqns}   in the
entire turbulent boundary layer is:
\BSE\label{FormalSol}
\BEA\label{FormalSol-1}
 W^+ \!\! &=& \! 2\,W^+_{xx} =4\,W^+_{yy} = 4\,W^+_{zz} \,, \\
\label{FormalSol-3}
W^+_{xy} \!\! &=&\! -\frac{3\,\gamma^+\,W^+}{2\,S^+}\,, \quad {S^+}^{2}\! = 15\,
\gamma^+\! \( \~\Gamma^+ +3\, \~\gamma^+ \Sb{RI}\) .~~~~~~
 \EEA
\ESE
The last equation comes from the solvability condition for the
system of Eqs. \Ref{MainEqns}: $\mbox{Det}= 0$. Introduce:
 \BE 
\label{v-intro}
 v^+ \=  \sqrt{W^+}\,, \ \, v^+_{\ast} \= \sqrt{W^+_{\ast}}\,, \
\, r
 \= 1 +\frac{{\~a}^{\,2}}{3\,\~b \Sb{RI}\,v^+_{\ast}\,y^+}\ . %
\EE Then
\BE \label{2FSol-inv}
W^+_{xy} = -\frac{W^+}{2}\,\sqrt{\frac{b}{2\,\~b\, r}}\ ,\quad 
S^+\!\! = \frac{3\,v^+ }{y^+}\sqrt{2\,b\,\~b\, r}\,,~~~~~~ \EE
and  Eq. \Ref{MeanMom} transforms into:
\BE\label{QuadraticEq}
{v^+}^{2} +  12\,\~b\,r  \,{v^+}/ y^+ -P^+\,\sqrt{ {8\,\~b\,r}/{b} }
= 0\ . \EE
This is just a quadratic equation for $v^+ = \sqrt{W^+}$ with a
unique positive solution:
\BE\label{vSolInf}
v^+ = \sqrt{P^+\sqrt{\frac{8\,\~b}{b}\,r\,}
 +\(\frac{6\,\~b}{y^+}\,r\)^2\,} -\frac{6\,\~b}{y^+}\,r\ .
\EE

 \textbf{ Comparison of the model and simulations}: Clearly, a
model with only 4 (or 5 if $C\Sb{IP}$ is counted in) fit parameters
cannot fit perfectly the profiles of all the physical quantities
that can be measured. Therefore the actual value of the only
remaining parameter $\~a$ should be determined with a choice of the
characteristics of turbulent boundary layers that we desire to
describe best. Foremost in any modeling should be the mean velocity
profile $V^+$ which is of crucial importance in a wide variety of
transport phenomena.   Therefore we chose the  value of $\~a$
 from  the best fit of $V^+(y^+)$ in the (quasi)
straight logarithmic region $ 30< y^+ < 350$:  
\BE\label{a}
 \widetilde{a} = 7.1\ .
 \EE
The resulting  mean velocity profile $ V^+(y^+)= \int_{0}^{y^+}\!\!\!
S^+(\xi) \, d\,\xi$,  in which $ S^+(\xi)$ is given by
 \Ref{2FSol-inv},  is shown in Fig.~\ref{f:1} by a solid line for
Re$_\tau=2000$. The dotted line represents data taken from direct
numerical simulations \cite{DNS}, for the same Re$_\tau$. There is
no significant difference between these plots in the viscous
sublayer, buffer and outer layers, where $\yp\lesssim 800$ i.e. in
about 40\% of the channel half-width $L^+=\mbox{Re}_\tau$. This
robustness of the mean velocity profile  is a consequence of the
fact that $V^+(y^+)$ is an integral of the mean shear $S^+$ which is
described very well both in the viscous and the outer layers.

\begin{figure}
  \centering
   \includegraphics[width=0.47 \textwidth]{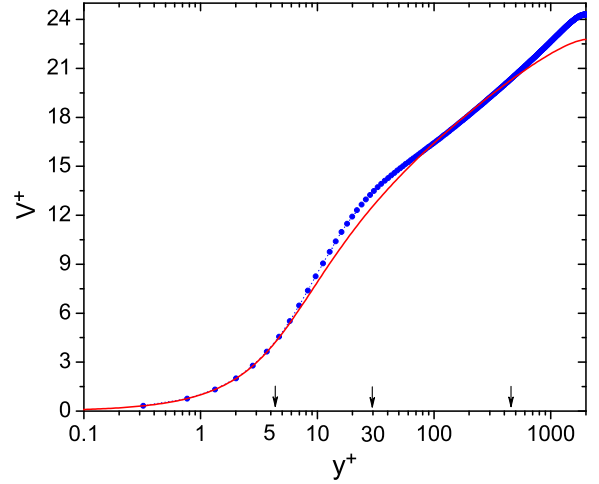}
  \caption{\label{f:1}    Mean velocity profiles $V^+(y^+)$: The
  dotted line reproduces the results of direct numerical simulations
  \cite{DNS} for Re$_\tau=2000$, the solid line  is the analytical
  prediction of the model with $\widetilde{a} = 7.1$ and the
  values~\Ref{all-b} of $b$, $b\Sb{RI}$ and  $\~b\Sb{Ri}$.}
\end{figure}

Notice that our model does not describe the upward deviation from
the log-low which is observed near the mid-channel (of about a few 
units in $V^+$).  We consider this minor disagreement as an
acceptable price for the simplicity of the model which neglects the
spatial energy transport term toward the centerline of the channel.
This transport is the only reason for some turbulent activity near
the centerline where both the Reynolds stress $W_{xy}$ and   $S $
vanish due to symmetry. Just at the center line the source term in
our energy equation, $ -2 S W_{xy}$, is zero, and the missing energy
transport term is felt.

\begin{figure}
  \centering
  \includegraphics[width=0.45 \textwidth]{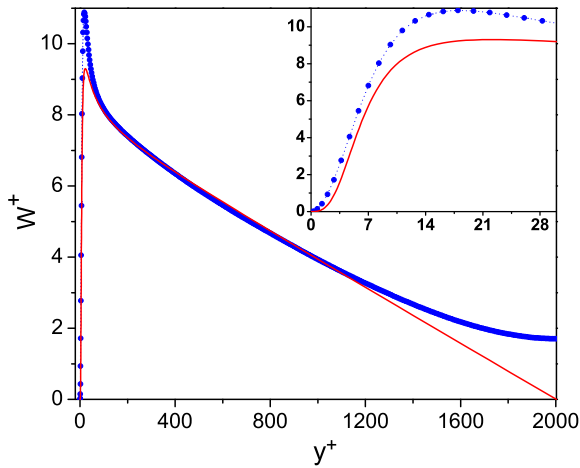}
  \caption{\label{f:2} The trace of the Reynolds-stress tensor
  (twice the total kinetic energy densiy): The dotted blue line
  reproduces the results of direct numerical simulations~\cite{DNS}
  at Re$_\tau=2000$,  the solid red line is analytical prediction
  with for $\widetilde{a} = 7.1$. The insert shows the buffer layer
  behavior in more detail. Notice, that there is no plateau in
  these plots, meaning that these values
of Re$_\tau$ are not large enough to have a true scale-invariant
log-law region.}
\end{figure}
The plots in Fig.~\ref{f:1} have a reasonably straight logarithmic
region from $\yp \approx 30$ to $\yp \approx 350$. On the other
hand, the Reynolds stress profile at the same Re$_\tau=2000$ shown in
Fig.~\ref{f:2},  has no flat region  at all. Such a flat region is
expected in the true asymptotic regime of Re$_\tau\to\infty$, where
$\Wp=-1$.  Therefore if one plots  the model profiles $V^+$ at
different Re$_\tau$ and fits them  by log-linear profiles 
\BE\label{karm}
 V^+(y^+)
= \kappa^{-1}\textrm{ln}(y^+)+ B\,, 
\EE
 one can get a Re$_\tau$-dependence of the ``effective" intercept
$B\Rightarrow B({\rm Re}_\tau)$ in the \vK log-law~\Ref{karm}.  We
think that this explains why measured values of the intercept depend
on the Reynolds number and on the flow geometry (channel vs. pipe):
both in direct numerical simulations and in physical experiments one
usually does not reach high enough values of Re$_\tau$.

   At this point $\~a$ and all three $b~$-parameters are already
chosen from the information about the mean velocity profile and
the $y^+$-independent values of the Reynolds stress tensor in the
lag-law region in the limit Re$_\tau\to \infty$.  In Figs.~\ref{f:2}
and \ref{f:3}  we show
 the predictions of the model for the kinetic energy profile
  $W^+$,  and the Reynolds stress profile (red solid lines). The result
  pertain to the entire turbulent boundary layer for Re$_\tau$=2000,
  without any additional fit parameters.  In the same figures we show
the results of direct numerical simulation ~\cite{DNS} (blue dotted lines).
  The excellent agreement in the entire turbulent boundary layer is a
  strong indication that the proposed model summarizes the
  important physical ingredients of the problem.

\begin{figure}
\centering
  \includegraphics[width=0.48  \textwidth]{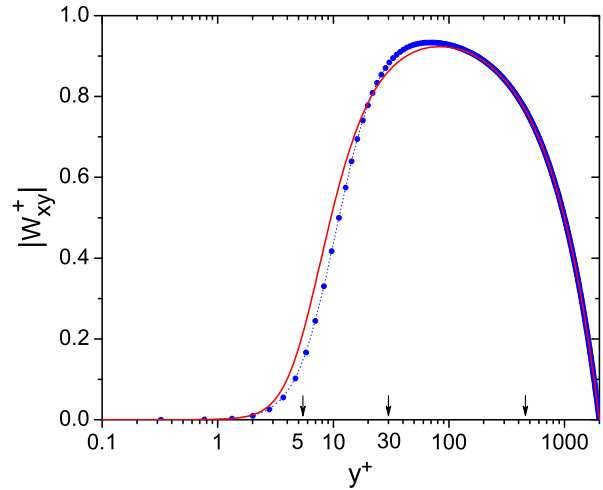}
  \caption{ \label{f:3} The Reynolds stress $ |W^+_{xy}|$ :
   The dotted blue
  line represents results from direct numerical simulations
  ~\cite{DNS} for Re$_\tau=2000$,   the solid red line  represents
  the prediction of  the analytical model with $\widetilde{a} =
  7.1$. There is no pronounced log-law region where a plateau
  $|W_{xy}^+|=1$ is expected.  Instead $|W_{xy}^+|$ reaches only the
  value $0.85$ around $y^+\!\!\approx 50$. This means that in this
  region the viscous transport is still essential, but the total
  momentum flux is still below its (dimensionless) maximum value:
  max$\(\P^+\) = 1$. } 
\end{figure}

\textbf{Summary}: In summary, we presented an analytic model of the
physics of wall-bounded turbulence in a Newtonian fluid, based
entirely on the balance of energy and momentum fluxes with the
production and the dissipation. The model has one blatant
simplification which is the neglect of the spatial energy fluxes.
The gain is enormous -- we get a local model that can be solved
analytically to find the profiles of the mean velocity, the
turbulent fluctuations and the Reyonlds stress as a function of the
distance from the wall, with all the boundary layer represented
without re-fitting in the various regions. The reason of the success
of this simple model is that we have learned how to compensate the
neglect of the spatial flux in the buffer and the viscous region by
a decrease in the dissipative terms. In future work we will
demonstrate the utility of this simple model in a variety of
important wall-bounded flows.

\textbf{Acknowledements:}   This research is supported by the
US-Israel Binational Science Foundation.

\end{document}